\documentstyle[12pt]{article}
\newcommand{\LI}{\hbox to\hsize}
\newcommand{\LLI}[1]{\LI{#1\hss}} \newcommand{\RLI}[1]{\LI{\hss#1}}

\newcommand{\PM}[1]%
{\mbox{$m_{\rm #1}$}} 
\newcommand{\BEQ}{\begin{equation}}
\newcommand{\EEQ}{\end{equation}}

\newcommand{\DEG}[1]{\mbox{$ #1^{\rm o}$}}
\newcommand{\ack}{\LLI{\Large{\bf Acknowledgement}}
\vspace*{2mm}
\par
\noindent
\nopagebreak
This research has been supported in part by the U.S. Department of Energy
under Grant \#~DE--FG--02--85ER40211} 
\newcommand{\gdskd}{\vskip10mm
\begin{center}
G. Domokos and S. Kovesi--Domokos \footnote{E--MAIL: SKD@JHUP4.PHA.JHU.EDU}
\\
The Henry A. Rowland Department of Physics and Astronomy \\
The Johns Hopkins University \\
Baltimore, MD 21218
\end{center}
\vskip10mm }
\newcommand{\incircle}[1]{\mbox{{\hbox{$\bigcirc$}\kern-0.7em
\lower0.05ex\hbox{\mbox{{\scriptsize\rm #1}}}}}}
\newcommand{\ETC}{\mbox{\em etc.\/ }}
\newcommand{\VIZ}{\mbox{\em viz.\/ }}
\newcommand{\CF}{\mbox{\em cf.\/ }}
\newcommand{\IE}{\mbox{\em i.e. \/}}

\newcommand{\EG}{\mbox{\em e.g.\/ }}

\newcommand{\mathfig}[4]{%
\begin{figure}[#4] 
\vspace*{6in}
\centerline{\hbox to 14in 
 {\hskip5in {\special{eps:#2 x=14in}\hfil}}} 
\vspace*{-1.5in} 
\caption{#3}
\label{#1}
\end{figure}\vskip0.5in} 
\begin{document}
\RLI{JHU--TIPAC 940009}
\begin{center}
{\Large\bf UNUSUAL EVENTS IN NEUTRINO TELESCOPES: SIGNATURES
OF NEW PHYSICS}\footnote{Invited talk given at the International Workshop on
Neutrino Telescopes, Venice 1994; to be published in the
Proceedings.}
\end{center}
\gdskd
\vspace{1.5cm}

{\small A class of physical phenomena outside the framework of the
perturbatively treated Standard Model of electroweak and strong
interactions gives rise to characteristic signatures in neutrino telescopes.
In essence, the signature is a large energy deposition in the neighborhood
of the telescope, giving rise to large and concentrated Cherenkov
light emission, and in some cases, to energetic muon bundles.}
\vskip1.5cm
\section{Introduction}
Despite the fact that the Standard Model of electroweak and strong
interactions is in a very good agreement with presently available
experimental data, the general situation concerning the model is
unsatisfactory. There are several reasons for making such a
statement. Here are a few.
\begin{itemize}
\item One has very little understanding of the mechanism of
the breaking of the electroweak symmetry: the Higgs boson
has yet to be found. Moreover, there are strong reasons to believe
that the theory based on an elementary Higgs field is internally
inconsistent.
\item We  understand very little about the structure of the
theory outside the framework of perturbation theory. One is
virtually certain that perturbation theory is badly divergent,
hence it is useless
in the strong coupling region. Consequently, our understanding of
important phenomena, like quark confinement is extremely
limited. (Lattice approximations do contain a hint that quantum
chromodynamics indeed confines quarks and gluons. However,
quantitative results are still unavailable, in part due to the enormity of
the computational problem and, in part, due to theoretical questions
connected with the continuum limit of the lattice approximation.
\item There are some interesting, non--perturbative phenomena
within the framework of the electroweak theory, such as
the violation of the {\em sum} of baryon and lepton numbers,
as predicted by 't Hooft, ref.~\cite{thooft}, which have
received much attention lately. In particular, Ringwald and
his collaborators conjectured that if such a process is
accompanied by a multiple production of gauge bosons
($Z \quad {\rm and} \quad W$), one may reach observable levels of cross
sections, see, \EG \cite{ringwald} for a recent review. If observed,
such phenomena would shed light on the properties of the ground state
of the theory (the vacuum) and, hence, it would be very interesting
to observe them.

However, it is to be emphasized that, so far, the
calculations have been
extremely unreliable; in fact, one's knowledge of  the magnitude
of the cross sections involved is both meager and shaky.
\end{itemize}

In addition to the reasons just mentioned,
one feels uneasy about a theory which
contains {\em nineteen} input parameters. (If neutrinos are
massive, the number of input parameters is  larger still.) Despite
the various schemes proposed (grand unified theories,
composite models, strings \ETC ) to remedy this situation,
very little real progress has been made during the last decade.
This is due, in part,
to the fact that one cannot make controllable approximations to the
theories just mentioned.

There is little doubt that observing a signature of any of the phenomena
alluded to above would be of great importance and it would
contribute significantly to our understanding of the physics
at energy scales above the currently reachable ones (of the order of
100 GeV at the level of quarks and
leptons).

In view of theoretical difficulties just mentioned,
we'll take a cautious approach: we'll try to extract some salient
features of the unreliable calculations, in the hope that
those are sufficiently robust and are largely independent
of the details of the models. Then, we'll investigate the
measurable consequences these phenomena in a neutrino
telescope.
\vskip1.5cm

\section{Common Features of Some Post Standard Model Scenarios}

Most scenarios conjectured to step beyond the Standard Model
have an important feature in common, \VIZ an excess production
of hadrons under unexpected circumstances. (Most supersymmetric
models do not belong to this category.) Let us illustrate this
point on two examples mentioned earlier and on one not yet mentioned.
\begin{enumerate}
\item {\em Preon models\/.} Even though most proposed preon models have been
less than spectacularly successful so far,
they may  constitute a reasonable step towards reducing the number of
arbitrary input parameters in the Standard Model.
Typically in any economical preon model,
quarks and leptons share at least some preon constituents. As a consequence,
at energies exceeding the characteristic scale of the model (say, the
energy at which a hypothetical metacolor gauge theory enters its strong
coupling regime), in a sense, both leptons and quarks become ``schizophrenic'':
in particular, lepton--quark interactions begin to produce energetic hadrons
in the projectile fragmentation regime. Scenarios of this kind and their
consequences in UHE neutrino physics have been discussed previously,
\CF \cite{domonuss, our88, steve} and references quoted therein.
It is rather hard to estimate the hadron multiplicity in the initial
interaction; however, if QCD can be used as a guide, one expects that
about half of the initial energy will go into a hard neutrino or
charged lepton and the other half into about 20 or so quark pairs,
\CF \cite{steve} and references quoted there. Thus the total
hadronic multiplicity is about 20 or so, carrying about half of the
primary energy.
\item {\em Multiple production of weak gauge bosons\/.}
Strictly speaking, this process is {\em within} the framework of the
Standard Model; however, it is definitely a non--perturbative phenomenon
and thus, it is, in a sense, ``new physics''. Due to the fact that
both W and Z decay into hadrons roughly 70\% of the time, if
multiple W/Z production takes places in a neutrino interaction, one is
likely to see, on the average, a large number of hadrons in the initial
interaction. As a rough estimate, let us assume that there are 20 weak gauge
bosons produced, see \EG \cite{ringwald}. On the average then, about 14
of them decay hadronically; the average multiplicity is of the order of 10,
mostly light mesons and a negligible number of baryon pairs. Thus, the
initial energy is expected to be distributed among about 200 mesons (not
counting the hadronic decay modes of the $\tau$) and some ten or so
hard leptons, depending on the (still ill--understood) details of the
process.
\end{enumerate}

In both the multi--W production and the compositeness scenarios
 it is essential to concentrate on $\nu$ induced reactions: in
any other type of reaction (with the possible exception of a $\gamma \gamma$
collider) the backgrounds are unacceptably high.
In both cases, the essential observation is that the estimated cross
sections are in the $\mu$b range or somewhat smaller; however, several orders
of magnitude larger than the neutrino cross sections given by perturbative
calculations within the framework of the standard model.

Let us observe that for cross sections of this magnitude and for CMS
energies in the multi--TeV range, the effects of nuclear structure
(surface absorption, \ETC) are entirely negligible. Consequently, the
target appears as a gas of nucleons to the projectile; hence the
interaction mfp is independent of A.
Numerically, one gets:
\BEQ
\lambda [{\rm g/cm^2}] \approx \frac{1670}{\sigma [{\rm mb}]}
\label{eq:mfp}
\EEQ

It follows that, according to eq.~\ref{eq:mfp}, a mfp of
 4,000 mwe ($\approx 4\times 10^{5} {\rm
g/cm^2}$) corresponds to a cross section of about 4 $\mu$b.
Hence, a neutrino incident nearly vertically will produce, on average,
the first interaction close to the typical neutrino telescope (DUMAND II
or NESTOR). Lower cross sections can be observed at higher zenith angles.
Details depend on the density profile surrounding the detector
and we shall not discuss this question any further at this time.
\vskip1.5cm

\section{The Qualitative Appearence of the Underwater Cascade}

While details differ in the first two scenarios outlined in the previous
Section, in both cases one produces a number of mesons going forward in
the CMS --- a phenomenon not expected in neutrino induced rections
within the framework of the Standard Model and perturbation theory.
For all practical purposes, there are no nucleon pairs produced in the
initial interaction; most of the mesons produced (about 90 \% or so)
are light ($\pi$, K).

At the relevant energies (several TeV in the LAB system), the
hadronic interaction mfp is of the order of 50 ${\rm g/cm}^2$. In water, this
corresponds to a distance of about 50 cm. This is to be compared with a
typical charged meson decay mfp which is of the order
of $ \gamma \times 8{\rm m}$.
(Here, $\gamma$ stands for the Lorentz factor of the meson in question.)
Hence, {\em light charged mesons do not decay\/}: there are no
delayed muons in the cascade. The only muons to be found are ``prompt''
ones, coming from either the primary interaction (in the first scenario
in the previous Section) or (mostly) from the deacy of the weak
gauge bosons in the second one. (An additional source of prompt muons
is the production of mesons containing c and b quarks: it is not clear
at present what fraction of those mesons will be; however, one
does not expect it to be very high.)

Neutral pions (about $1/3^d$ of all mesons produced) will decay rather
than interact: as a consequence, there will be a very substantial
electromagnetic (EM) component in the cascade. Most of the EM
component will originate from $\pi^0$ (and $\eta^0$) decay. In the second
scenario discussed in the previous Section, there is a small
(at the $\approx 10 \%$ level) prompt EM component originating from the
decay of the weak gauge bosons into electrons (and $\nu_e$ in the case
of the charged gauge bosons). We do not believe that one can
realistically expect a distinction between the prompt and delayed
parts of the EM components of the cascade within this
century. Hence, we shall concentrate on the
EM component arising from $\pi^0 \quad {\rm and} \quad \eta^0$
decay from now on.

In what follows, we are going to present a quantitative picture of the
longitudinal development, based on an approximate casacade theory as
outlined in ref.~\cite{pylos93}.
\vskip1.5cm

\section{Calculation of the Cascade Development}

Due to the large theoretical uncertainties in the primary interaction,
we can simplify matters considerably in the casacade calculation. Some of the
simplifications introduced have been discussed previously,
\CF ref.~\cite{pylos93}. Below, we briefly summarize the simplifications.
\begin{itemize}
\item The  majority of mesons produced is a pion; we neglect the
production of mesons containing s, c, b and t quarks. (This
simplification introduces an error of the order of 15 \%.)
\item The mesons $\pi^{\pm}$ do not decay, the interaction rate
of any $\pi^{0}$ is negligible compared to its decay. Hence,
by charge symmetry, about $2/3^d$ of the mesons participate
in the cascade, the remaining ones feed the electromagnetic component.
\item There
are practically no initial baryons in the casacade and baryon
pair production is negligible: one can work with a single
component hadronic cascade.
\item Photoproduction of mesons is small: the photoproduction
cross section is about 1\% of the hadronic inelastic one.
Hence, the hadronic component develops autonomously,
with a negligible feedback from the electromagnetic one.
\item As a consequence, the electromagnetic component has a
a source (from the process $\pi^{0} \rightarrow \gamma \gamma$),
otherwise, it develops on its own.
\item Approximation A is adequate for both components since
we are mainly interested in the  high energy component
of the cascade.
\end{itemize}
We have shown in ref.~\cite{pylos93} that even atmospheric cascades
can be treated reasonably accurately (at  a
level of error about 35\% or so at the highest energies)
 under these simplifications. Due to the absence of nucleons, the treatment
should work  better in the present case.

With the simplifications mentioned above, the cascade theory
is a linear one to a high degree of accuracy. As a consequence,
it is not important to
specify the multiplicity in the initial interaction precisely;
multiplicities in the cascade scale linearly with the initial one.
(Linearity breaks down at the lowest energies considered:
low energy mesons produce fewer secondaries than high enrgy ones
and the cascade stops.)
Similarly, since, {\em on the average} the portion of the initial
neutrino energy going into meson production  is distributed
uniformly among the mesons, the average meson energy at the
beginning of the cascade is given by the simple formula:
\BEQ
E_{1} = \frac{\kappa E_{\nu}}{ \langle N_{0}\rangle },
\EEQ
where $\kappa$ stands for the inelasticity (we estimate it to be about
1/2) and $ \langle N_{0}\rangle $ is the average multiplicity in the first
interaction. Later on, we plot the evolution of a cascade by taking
 $\kappa E_{\nu} = 10^{17}{\rm eV}$ and (in order to study the
the effect of the nonlinearity at ``low'' energies) we
plotted the cases $ \langle N_{0}\rangle = 20$ and $ \langle N_{0}\rangle
 = 5$, respectively.

\subsection{The Hadronic Component}

We assume the validity of Feynman scaling. (The validity of this
approximation has been discussed in some detail in ref.~\cite{pylos93}.)
In the diffusion
approximation and neglecting decay, the cascade equation for the hadronic
component reads:
\BEQ
\frac{\partial H\left(E, x\right)}{\partial x} = - H\left( E, x\right)
+ \int_{E}^{\infty}\! \frac{dE'}{E}
F\left(\frac{E}{E'}\right) H\left(E', x\right)
\EEQ
Here $x$ stands for the thickness measured in units of the
hadronic interaction mfp. The fragmentation function, $F(z)$
is taken to be of the form:
\BEQ
F\left(z\right) = C z^{-1 +\epsilon}\left( 1-z\right)^3
\Theta\left( 1- z\right).
\EEQ
The normalization  constant, $C$ and the infrared regulator, $\epsilon$
are introduced in
order to satisfy the constraints:
\[ \int \! dz F =1
\]
and
\[ \int \!\frac{dz}{z} F = \langle N_{h}\rangle,
\]
$\langle N_{h}\rangle$ being the average multiplicity
in a hadronic interaction. (We chose $\langle N_{h}\rangle = 30$.)
Most results which follow are rather insensitive to the precise
value of $\epsilon$.

We solve this equation for a monochromatic initial spectrum in order to
be able to follow the longitudinal development of the cascade in some
detail. The initial condition is:
\BEQ
H\left( E, 0\right) =  \langle  N_{0}\rangle \delta
\left( E - \kappa E_{\nu}\right)
\EEQ

The solution is obtained by means of the iterative procedure
described in ref.~\cite{pylos93}. The evolution is cut off
when the CMS energy in the interactions drops below
$\sqrt{s} = 30$ GeV.

In the following Figures
we exhibit the result
for the initial energy and multiplicities described above.
We are interested in the integral spectrum of hadrons and of the
EM component.

Figure~\ref{long5}  displays the longitudinal
structure of the cascades for a low initial
multiplicity $N_{0}=5$. The two curves  correspond to
energies $E>10$ GeV (upper curve) and $E>100$ GeV (lower curve),
respectively.

\mathfig{long5}{/math/figures/bomban5.ps}{Longitudinal distribution
of hadrons; initial multiplicity, $N_{0}=5$}{t}

\newpage

As expected, the hadronic distribution
is quite narrow in $x$ and, similarly in real space. For
all practical purposes, the energetic hadrons are concentrated within
a space of about 6 m in water.
Figure~\ref{long20} contains the same information as the
previous one, but with an initial multiplicity, $N_{0}=20$.

\mathfig{long20}{/math/figures/bomban20.ps}{Longitudinal distribution
of hadrons; initial multiplicity, $N_{0}=20$}{t}

The shower profiles are quite similar for $E>10\, {\rm GeV}$,
provided one scales
the curves  appropriately.
Higher energies are more affected by a larger initial multiplicity,
due to the fact that, on the average, higher multiplicities
lower the initial emergy per particle in the primary interaction.
\subsection{The Electromagnetic Component}
This component is treated within the framework of Approximation A.
The electromagnetic cascade equations are inhomogeneous, the
source term for photons,  $S(E,\xi)$ being given by
\BEQ
S\left( E, \xi \right) = \frac{2}{3} H\left( 2E ,\rho \xi \right).
\EEQ

Here $\xi$ stands for the distance measured in units of
the radiation length ($X_{0}$) and $\rho$ is a conversion factor between
the hadronic interaction mfp and and $X_{0}$. Due to the fact
that water is very nearly incompressible, $\rho$ is a constant to
a very good approximation, $\rho \approx 0.4$.

Apart form the initial stages of the electromagnetic cascade,
the number of electrons and positrons is nearly the same as
that of the photons, with a slight photon excess due to the source
term.

The solution of the cascade equations with a source term is a staright
forward one, once the retarded Green function is found. The latter is best
determined by means of an iterative procedure, similar to the one
described in ref.~\cite{pylos93}; we shall not repeat the description here.

In the following we do not distinguish between the photon and
electron--positron components; within the accuracy of the calculation,
there is no significant difference between them.
In Fig.~\ref{electro} we plot the longitudinal evolution of the
electromagnetic cascade for particles of energy $E>10$ GeV. (A typical
water Cherenkov detector is nearly 100\% efficient above such an energy.)

\mathfig{electro}{c:/math/figures/elgamma.ps}{The longitudinal evolution
of the electron -- photon component of the cascade. The abscissa
is the absorber depth in units of the radiation length,
$x_{r}$.}{t}

We plotted the evolution for one primary energy and initial multiplicity
only: this is sufficient to illustrate the qualitative
features. (For a lower initial multiplicity, the shower maximum occurs
further down along the cascade, but the number of particle at
maximum is higher.) The overall longitudinal size
of the electromagnetic component is comparable to the hadronic one
(about 6 m); however, the electromagnetic component peaks at
 $\approx 4$ m after the hadronic one. It is worth investigating
whether the two separate peaks in energy deposition
are observable: if so, this would give rise to a well--recognizable
signature in neutrino telescopes of events of this type.

\section{Discussion}

The estimates described here suggest that neutrino telescopes may
play an important role in particle physics in discovering phenomena
beyond the Standard Model. The search strategy in a neutrino telescope
{\em with} directional capability (\EG NESTOR) is to scan the
range of zenith angles available and serach for large energy
deposition within the span of a few meters. Once such events
are found, knowing the direction of incidence of the primary
allows one to estimate the cross section of the primary
interaction. Even without being able to tell the direction of
incidence (\IE without the ability to distinguish between zenith
angles larger and smaller than \DEG{90}), one can gain useful
information: it is hard to think about processes in the cross section
range between a few $\mu$barns and the Standard Model neutrino
cross section with a large energy deposition in a small volume.

In the previous considerations it was implicitly assumed that
after the primary interaction, the evolution of the cascade
follows Standard Model physics. While such an assumption is
somewhat {\em ad hoc}, probably it can be justified by
arguing that at currently available energies at accelerators
the Standard Model is in excellent agreement with the data.
Assuming that the primary event creates a sufficiently
high number of secondaries, one expects that already the
first generation of secondaries will have insufficient energy
in order to deviate significantly in its behavior from the
Standard model.

Assuming optimistically that ``anomalous'' events of the
kind described here will be found, one can speculate whether
one will be able to distinguish between the scenarios sketched
before. While we cannot overemphasize  the large
theoretical uncertainties, one may, nevertheless, see a few
potential differences:
\begin{itemize}
\item In the multi -- W/Z production scheme, the initial interaction
tends to produce a substantial number of electrons and muons.
As a consequence, the electromagnetic component of the cascade
starts right at the primary interaction
rather than being separated by a few hadronic mfp--s
from it. Also, one expects
a muon multiplicity averaging perhaps three or four, appearing
in the form of (almost) collinear muon bundles.
\item By contrast, in the scheme of composite models
(equally beset by theoretical difficulties), due to the
fact that most incident neutrinos are expected to
be $\nu_{\mu}$, probably, one has {\em one} very energetic
muon emerging from the primary interaction and otherwise
hardly any other muons being present.
\end{itemize}

It is amusing (and perhaps relevant) to remark that the description of
at least
one event roughly fitting the characteristics of an anomalous
event outlined here has been published in the
literature, see~\cite{kamiokande}. It may be interesting to examine
the record of other underground detectors for the occurrence of similar
events.

Clearly, more work on this subject is needed --- both by theorists
and experimentalists: it is a very interesting challenge. This century
began by the discoveries of what we call now the fundamental elements
of modern physics --- the discovery  of
cosmic rays by Victor Hess was among
them. Not only did Hess' discovery lead to a number of important
discoveries in the emerging
science of elementary particle physics, it also contributed substantially
to our understanding of the Universe we are living in. It would be quite
interesting and pleasing if, by the turn of this century, cosmic
ray physics rose to a renewed prominence by the joint effort
of particle physicists and astrophysicists.
\ack \/.

We  thank  Milya
Bilenky, Christine Kourkoumelis, Al Mann and Leonidas Resvanis for
useful discussions.

Last but not least,
we thank Milla Baldo--Ceolin and her colleagues for organizing this
stimulating workshop.
\vskip5mm
\LLI{{\Large\bf References}} 

\end{document}